\documentclass[a4paper,twoside,10pt]{letter}
\usepackage{saj,graphicx,multicol,subeqnarray}

\def\papertype{Original Scientific Paper}

\def\degr{\hbox{$^\circ$}}
\def\arcmin{\hbox{$^\prime$}}
\def\arcsec{\hbox{$^{\prime\prime}$}}
\def\SNR{\mbox{{SNR~J0450--709}}}
\def\udc{524.354--77 : 524.722.3}
\begin{document}
\baselineskip=3.1truemm
\columnsep=.5truecm
\newenvironment{lefteqnarray}{\arraycolsep=0pt\begin{eqnarray}}
{\end{eqnarray}\protect\aftergroup\ignorespaces}
\newenvironment{lefteqnarray*}{\arraycolsep=0pt\begin{eqnarray*}}
{\end{eqnarray*}\protect\aftergroup\ignorespaces}
\newenvironment{leftsubeqnarray}{\arraycolsep=0pt\begin{subeqnarray}}
{\end{subeqnarray}\protect\aftergroup\ignorespaces}
%


\markboth{\eightrm MULTIFREQUENCY OBSERVATIONS OF ONE OF THE LARGEST LOCAL GROUP SNRs, LMC J0450--709}
{\eightrm K. O. {\v C}ajko, E. J. Crawford and M. D. Filipovi\'c}

{\ }

\publ

\type

{\ }


\title{MULTIFREQUENCY OBSERVATIONS OF ONE OF THE LARGEST SUPERNOVA REMNANTS IN THE LOCAL GROUP OF GALAXIES, LMC -- \SNR} 


\authors{K.O.~{\v C}ajko$^{1}$, E.J. Crawford $^2$, M.D.~Filipovi\'c$^2$}

\vskip3mm


\address{$^1$Faculty of Sciences, University of Novi Sad
\break Trg Dositeja Obradovi\'ca 4, 21000 Novi Sad, Serbia} 

\Email{tinacaj@gmail.com}

\address{$^2$School of Computing and Mathematics, University of Western 
Sydney\break Locked Bag 1797, Penrith South DC, NSW 1797, Australia}  

\Email{e.crawford@uws.edu.au, m.filipovic@uws.edu.au}


\dates{July 2009}{July 2009}


\summary{We present the results of new Australia Telescope Compact Array (ATCA) observations of one of the largest supernova remnants, \SNR, in the Local Group of galaxies. We found that this Large Magellanic Cloud (LMC) object exhibits a typical morphology of an old supernova remnant (SNR) with diameter D=102$\times$75$\pm$1~pc and radio spectral index $\alpha$=--0.43$\pm$0.06. Regions of high polarisation were detected with peak value of $\sim$40\%. }


\keywords{ISM: supernova remnants -- Magellanic Clouds -- Radio
  Continuum: ISM -- Polarisation -- ISM: individual objects -- SNR
  J0450--709}

\begin{multicols}{2}
{


\section{1. INTRODUCTION}

The Magellanic Clouds (MCs) are one of the most favorable places for investigations of objects such as supernova remnants (SNRs). Beside their position, close to the South Pole, they are found in one of the coldest parts of the radio sky allowing us to investigate and detect radio emission without interruptions from Galactic foreground radiation (Haynes et al.~1991). The MCs are located outside of the Galactic plane and thus the influence of dust, gas and stars are small, making detailed investigations of SNRs possible. The Large Magellanic Cloud (LMC) at a distance of 50~kpc (Hilditch et al.~2005), allows for analysis of the energetics of each remnant. 

SNRs are usually characterised by their strong and predominately non-thermal emission that they emit at radio wavelengths. They have typical spectral index of $\alpha\sim-0.5$ defined by $S\propto\nu^\alpha$. SNRs have very important influence on the interstellar material (ISM). Appearance of shell like filaments are very often perturbed by interaction with and non-homogeneous structure of the ISM. SNRs dictate behavior, structure, morphology and evolution of the ISM. But on the other side, the evolution of SNRs is dependent on the environment which surrounds them. 

Here, we report on new moderate resolution radio-continuum observations of \SNR, one of the largest SNRs in the Local Group of Galaxies. It was initially classified as SNR by Mathewson et al.~(1985), based on optical observations (H$\alpha$) and later, radio-continuum survey with MOlonglo Synthesis Telescope -- MOST. McGee et al.~(1972) named this source as MC\,11 in their 4800~MHz MCs catalogue, but didn't attempt to classify it. Clarke et al.~(1976) also noted \SNR\ in their 408~MHz MC4 catalogue. Wright et al.~(1994) catalogued \SNR\ as PMN~J0450--7050 based on their observations with Parkes-MIT-NRAO (PMN) surveys at 4850~MHz. Filipovi\'c et al.~(1995, 1996) added further confirmation with a set of radio-continuum observations (with Parkes telescope) over a wide frequency range. Filipovi\'c et al.~(1998a), using {\it ROSAT} All Sky Survey (RASS) observations, did not detect X-ray emission from \SNR. Neither the {\it ROSAT} PSPC and/or HRI observations covered the area of \SNR. Blair et al.~(2006) reported detection at far ultraviolet wavelengths based on FUSE (Far Ultraviolet Spectroscopic Explorer) satellite, and named SNR as D90401. As very faint object \SNR\ is detected at [C\textsc{iii}] and [O\textsc{vi}] wavelengths. 

The [S\textsc{ii}]/H$\alpha$ ratio is 0.7, which according to Mathewson et al.~(1985) satisfies one of  the selection criteria for the large diameter class of SNRs. H$\alpha$ emission from \SNR\ shows typical well evolved shell-like appearance. Filipovi\'c et al.~(1998b) presented Parkes low resolution multi-frequency analysis and estimated spectral index of $\alpha=-0.39\pm0.08$. Williams~et~al.~(2004) added high resolution XMM-Newton X-ray results. Most recently, Payne et al.~(2008) presented optical spectroscopy of a wide range of LMC SNRs including \SNR. They found an enhanced [S\textsc{ii}]/H$_\alpha$ ratio of 0.5 typical for SNRs.

\section{2. OBSERVATIONAL DATA}

We observed \SNR\ with the Australia Telescope Compact Array (ATCA) on 6$^\mathrm{th}$ April 1997, with an array configuration EW375, at wavelengths of 8640 and 4800~MHz ($\lambda$=3 and 6~cm). Baselines formed with the $6^\mathrm{th}$ ATCA antenna were excluded, as the other five antennas were arranged in a compact configuration. The observations were carried out in the so called ``snap-shot'' mode, totaling $\sim$1 hour of integration over a 12 hour period. Source PKS B1934-638 was used for primary calibration and source PKS~B0530-727 was used for secondary calibration. The \textsc{miriad} (Sault and Killeen~2006) and \textsc{karma} (Gooch~2006) software packages were used for reduction and analysis. More information on the observing procedure and other sources observed in this session can be found in Boji\v{c}i\'c~et~al.~(2007) and Crawford~et~al.~(2008a,b).

Images were prepared, cleaned and deconvolved using \textsc{miriad} tasks, using multifrequency synthesis (Sault and Wieringa~1994). The 4800~MHz image (Fig.~1) has a resolution of 21\arcsec$\times$19\arcsec\ and an estimated r.m.s. noise of 0.5~mJy/beam. As the remnant is larger than the primary beam response (5\arcmin) at 8640~MHz no reliable image could be prepared.

\section{3. RESULTS AND DISCUSSION}

The remnant has a shell like morphology centered at RA(J2000)=4$^h$50$^m$33.4$^s$, DEC(J2000)=--70\degr50\arcmin 43.5\arcsec\ with a diameter of 420\arcsec$\times$310\arcsec$\pm$5\arcsec\ (7\arcmin$\times$5.16\arcmin\ or 101.8$\times$75.2$\pm$1~pc), which is in agreement with optical diameter of 6.5\arcmin$\times$4.7\arcmin reported by Williams~et~al.~(2004). We note that it is elongated in north-south direction and that is ``clumpy'' along its western side. 

Flux density measurements were made at 4800~MHz resulting in value 0.448~Jy. New measurements were also made at 843~MHz (from the LMC MOST image), 1400~MHz and 8640~MHz (both from the mosaics presented by Filipovi\'c et al.~2009 and Hughes~et~al.~2007). Using values of flux densities obtained from observed frequencies in Table~1, a spectral index was plotted (Fig.~4) and estimated to be $\alpha=-0.43\pm0.06$, confirming the non-thermal nature of this object as the still dominant emission mechanism. However, our value is slightly ``flatter'' in comparison with typical and estimated value of $-0.5$ for SNRs (Mathewson et al.~1985). 

Uro{\v s}evi{\' c} and Pannuti (2005) showed models for two cases when SNRs could produce significant amount of thermal emission. Further on, they discuss the contribution of that thermal emission to the radio-continuum spectral index make-up of SNR. Uro{\v s}evi{\' c} and Pannuti (2005) derived flatter empirical $\Sigma-D$ relation which is in a good agreement with previous modified theoretical relations. Discrepancies between theoretically derived and empirically measured $\Sigma-D$ relations may be partially explained by taking into account thermal emission at radio frequencies from SNRs at particular evolutionary stages and located in particular environments. In the case of \SNR, this may indicate an older age for the remnant where contribution of thermal component could be significant, similar to the example given in Uro{\v s}evi{\' c} et al.~(2007) and elaborated on in Oni{\' c} and  Uro{\v s}evi{\' c} (2008). Also, \SNR\ is most likely expanding in a denser and warmer medium of n~$\sim1-10\,$cm$^{-3}$. It is well known that at higher frequencies SNRs flux density decreases which (for more details see Fig.~4). 

We note that the point at 408~MHz (Table~1; Fig.~4) vary slightly off the line of best fit. This is most likely due to an older data (1970-ties) processing. Particularly in this case it may overestimates flux density due to a clean bias effect. We estimate that the combined flux density errors from all radio images used in this study, are less than 10\%\ at each given frequency. 

Linear polarisation image for \SNR\ at 4800~MHz is shown on Fig.~3. Regions of fractional polarisation are quite strong. They are designated with polarisation vectors located at north-west side of the shell. Linear polarisation images for each frequency were created using \textit{Q} and \textit{U} parameters. While we do not have any measurements at 8640~MHz, the 4800~MHz image reveals some strong linear polarisation. Without reliable polarisation measurements at another wavelength, we could not determine the Faraday rotation. The mean fractional polarisation at 4800~MHz was calculated using flux density and polarisation:
\begin{equation}
P=\frac{\sqrt{S_{Q}^{2}+S_{U}^{2}}}{S_{I}}\cdot 100\%
\end{equation}
\noindent where $S_{Q}, S_{U}$ and $S_{I}$ are integrated intensities for \textit{Q}, \textit{U} and \textit{I} Stokes parameters. Our estimated peak value is $P\cong 40\%$. Along the shell there is a pocket of uniform polarisation possibly indicating varied dynamics along the shell. This unordered polarisation is consistent with the appearance of an older SNR.

}

\end{multicols}

\vskip5mm

\centerline{{\bf Table 1.} Integrated Flux Density of \SNR.}
\vskip2mm
\centerline{\begin{tabular}{|c|c|c|c|c|c|}
\hline\noalign{\smallskip}
 &S$_\mathrm{I}$ (0.408~GHz)&S$_\mathrm{I}$ (0.843~GHz)&S$_\mathrm{I}$ (1.4~GHz)&S$_\mathrm{I}$ (4.8~GHz)&S$_\mathrm{I}$ (8.64~GHz) \\
 \hline\noalign{\smallskip}
 \SNR          &1470 mJy& 837.3 mJy & 643.7 mJy & 448 mJy & 360 mJy \\
 \hline\noalign{\smallskip}
 Reference     &Clarke  & Mills     & This      &   This  & This \\
               &et al. 1976 &et al. 1984& Work  &Work     & Work \\
  \hline\noalign{\smallskip}
\end{tabular}}

\vspace{0.5cm}

\centerline{\includegraphics[width=.6\textwidth]{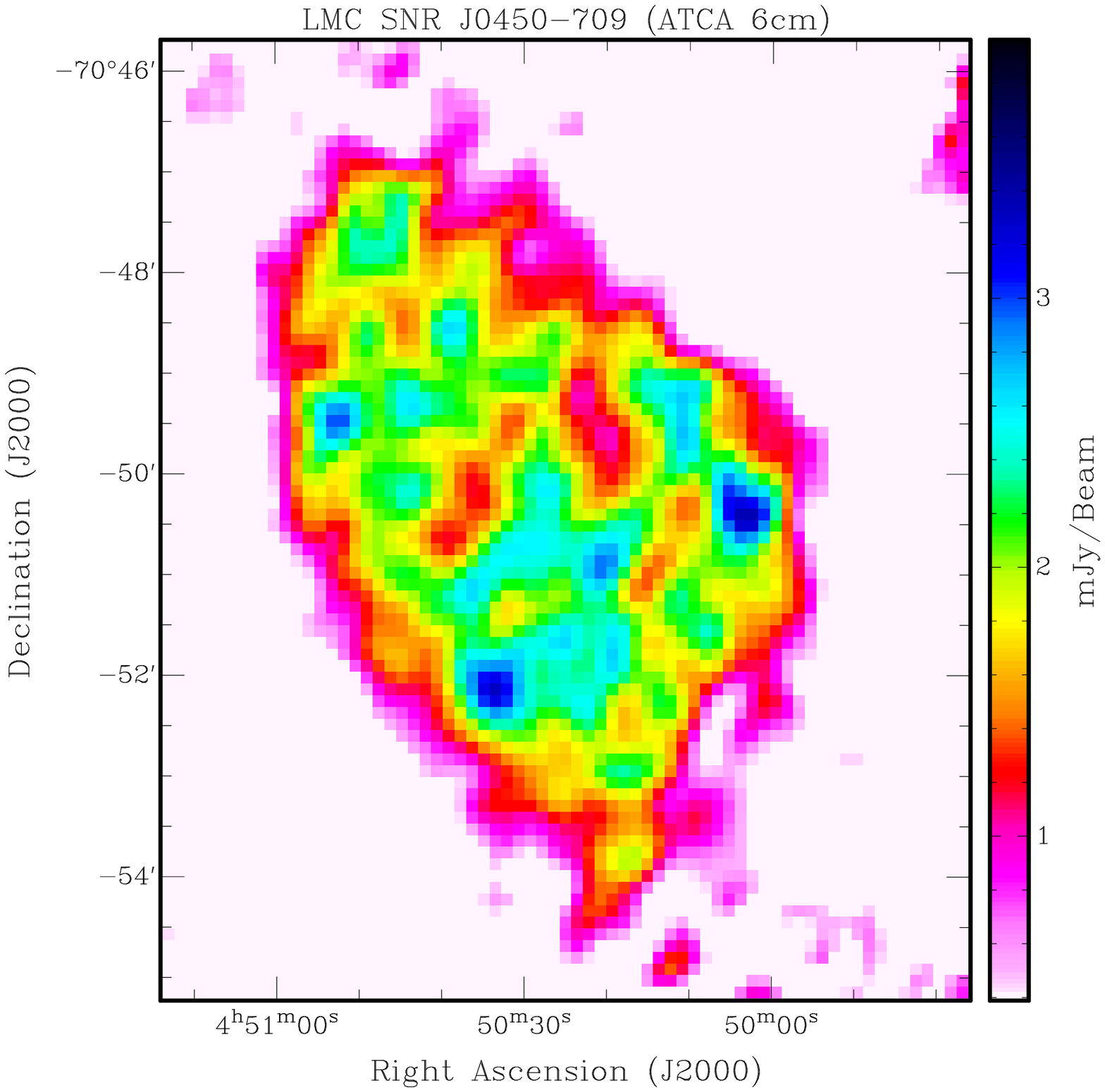}}
\figurecaption{1.}{ATCA observations of \SNR\ at 4800~MHz (6~cm). The side bar quantifies the pixel map and its units are mJy/beam.}

\centerline{\includegraphics[width=0.55\textwidth,angle=-90]{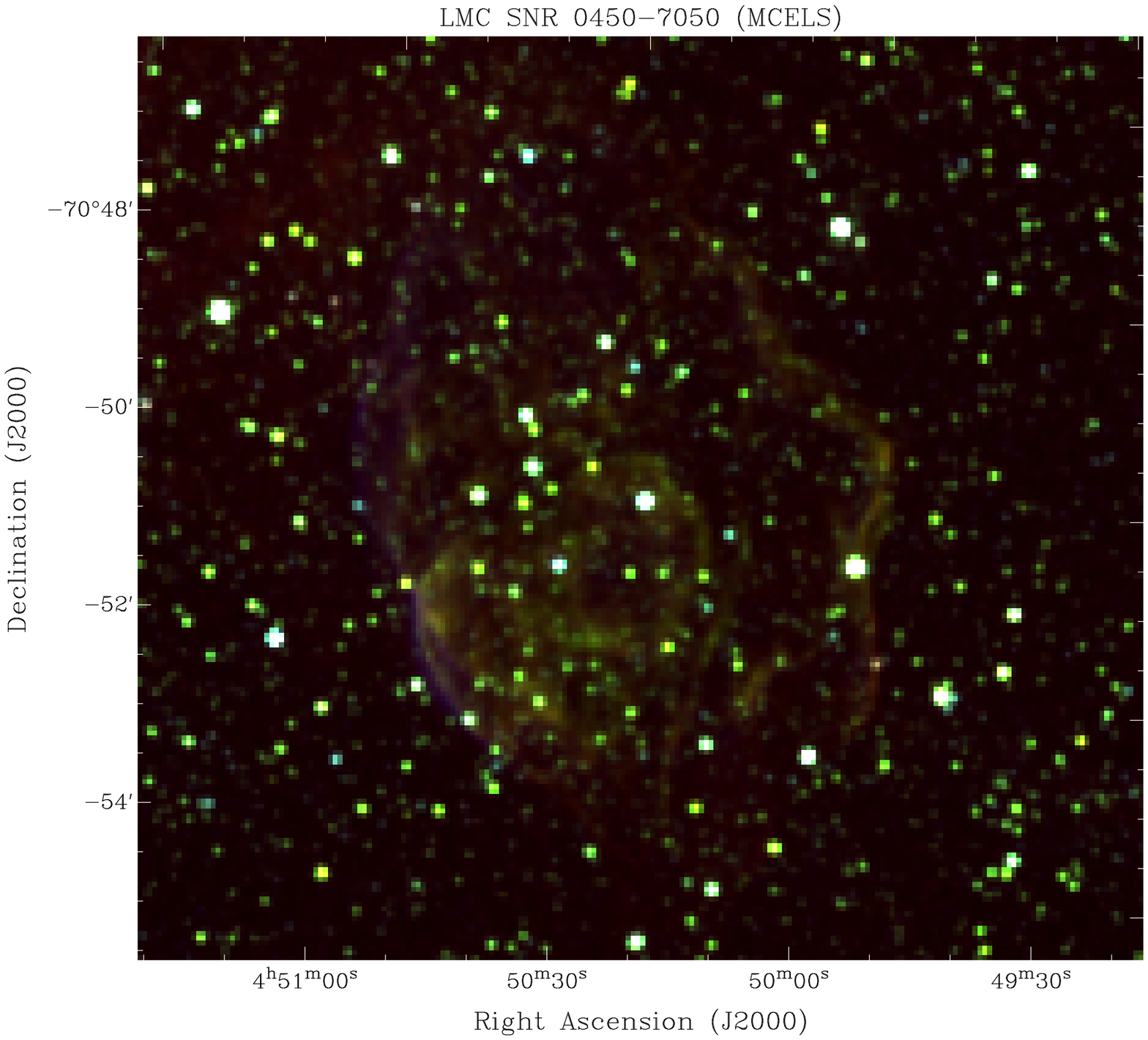}}
\figurecaption{2.}{MCELS composite optical image \textrm{(RGB =H$_\alpha$,[S\textsc{ii}],[O\textsc{iii}])} of \SNR}

\centerline{\includegraphics[width=0.55\textwidth,angle=-90]{fig-3}}
\figurecaption{3.}{ATCA observations of \SNR\ at 4800~MHz (6~cm). The blue circle in the lower left corner represents the synthesised beam-width of 21\,\arcsec$\times$19\,\arcsec, and the blue line below the circle is a polarisation vector of 100\%. The sidebar quantifies the pixel map and its units are Jy/beam.}

\centerline{\includegraphics[width=.55\textwidth,angle=-90]{fig-4} }
\figurecaption{4.}{Radio-continuum spectrum of \SNR.}

\begin{multicols}{2}

{

\section{4. CONCLUSION}

\vskip-1mm

We analysed one of the largest SNRs in the Local Group of Galaxies -- \SNR. Here, are presented new radio-continuum observations of this SNR together with the multi-frequency analysis. From these new observations, we found SNR diameter of \mbox{101.8$\times$75.2$\pm$1~pc,} spectral index of \mbox{$\alpha$=--0.43$\pm$0.06} and relatively strong level of linear polarisation with peak value of $\sim40\%$. We conclude that these are all indicators of an older SNR.


\acknowledgements{We used the {\sc karma} software package developed by the ATNF. The Australia Telescope Compact Array is part of the Australia Telescope which is funded by the Commonwealth of Australia for operation as a National Facility managed by CSIRO. We thank the Magellanic Clouds Emission Line Survey (MCELS) team for access to the optical images. }

\vskip.6cm


\references

\vskip-1mm

Blair, W.~P., Ghavamian, P., Sankrit, R., Danforth, C.~W.: 2006, 
\journal{Astron. Astrophys. Suppl. Ser.}, \vol{165}, 480. 

Boji{\v c}i{\'c}, I.~S., Filipovi{\'c}, M.~D., Parker, Q.~A., Payne, J.~L., Jones, P.~A., Reid, W., Kawamura, A., Fukui, Y.: 2007, \journal{Mon. Not. R. Astron. Soc.}, \vol{378}, 1237.

Clarke, J.~N., Little, A.~G., Mills, B.~Y.: 1976, \journal{Aust. J. Phys. 
Astrophys. Suppl.}, \vol{40}, 1. 

Crawford E.~J., Filipovi{\'c}, M.~D. and Payne, J.~L.: 2008 \journal{SAJ}, 
\vol{176}, 59. 

Crawford E.~J., Filipovi{\'c}, M.~D., De Horta, A.~Y., Stootman, F.~H., Payne J.~L.: 2008, \journal{SAJ}, \vol{177}, 61.

Filipovi\'c, M.~D., Haynes, R.~F., White, G.~L., Jones, P.~A., Klein, U., Wielebinski, R.: 1995, \journal{Astrophysical Journal Supplement Series}, \vol{111}, 331.

Filipovi\'c, M.~D., White, G.~L., Haynes, R.~F., Jones, P.~A., Meinert, D., Wielebinski, R., Klein, U.: 1996, \journal{Astrophysical Journal Supplement Series}, \vol{120}, 77. 

Filipovi\'c, M.~D., Pietsch, W., Haynes, R.~F., White, G.~L., Jones, P.~A., Wielebinski, R., Klein, U., Dennerl, K., Kahabka, P., Lazendi{\'c}, J.~S.: 1998a, \journal{Astron. Astrophys. Journ. Suppl. Ser.}. \vol{127}, 119.

Filipovi\'c, M.~D., Haynes, R.~F., White, G.~L., Jones, P.~A.: 1998b, \journal{Astron. Astrophys. Journ. Suppl. Ser.}. \vol{130}, 421.

Filipovi\'c M. D., Crawford E. J., Hughes A., Leverenz H., de 
Horta A. Y., Payne J. L., Staveley-Smith L., Dickel J. R., 
Stootman F. H., White G. L.: 2009, in van Loon J. T., 
Oliveira J. M., eds, \journal{IAU Symposium Vol. 256 of IAU Sym- 
posium}, pp PDFÐ8. 

Gooch, R.: 2006, Karma Users Manual, ATNF, Sydney.

Haynes et al.,: 1991, \journal{Astronomy and Astrophysics}, \vol{252}, 456.

Hilditch, R. W., Howarth, I. D., Harries, T. J.: 2005, \journal{Mon. Not. R. Astron. Soc.}, \vol{357}, 304. 

Hughes, A., Staveley-Smith, L., Kim, S., Wolleben, M., Filipovi{\'c}, M. D.: 2007,  \journal{Mon. Not. R. Astron. Soc.}, \vol{382}, 543.

Mathewson, D.~S., Ford, V.~L., Tuohy, I.~R., Mills, B.~Y., Turtle A.~J., Helfand, D.~J.: 1985, \journal{Astrophys. J. Suppl. Series}, \vol{58}, 197.

McGee, R.~X., Brooks, J.~W., Batchelor, R.~A.: 1972, \journal{Aust. J. Phys}, \vol{25}, 581.

Mills, B.~Y., Turtle, A.~J., Little, A.~G.,  Durdin, J.~M.: 1984, \journal{Aust. J. Phys.}, \vol{37}, 321.

Oni{\' c}, D., Uro{\v s}evi{\' c} D.: 2008, \journal{SAJ}, \vol{177}, 67.

Payne, J.~L., White, G.~L., Filipovi{\'c}, M.~D.: 2008, \journal{Mon. Not. R. Astron. Soc.}, \vol{383}, 1175.

Sault, R., Killeen, N.: 2006, Miriad Users Guide, ATNF, Sydney.

Sault, R.~J., Wieringa, M.~H.: 1994, \journal{Astron. Astrophys. Suppl. Ser.}, \vol{108}, 585.

Uro{\v s}evi{\' c}, D., Pannuti, T.~G.: 2005, \journal{Astroparticle Physics}, \vol{23}, 577.

Uro{\v s}evi{\' c}, D., Pannuti, T.~G., Leahy, D.: 2007, \journal{ApJL}, \vol{655}, L41.

Williams, R.~M., Chu, Y.-H., Dickel, J.~R., Gruendl, R.A., Shelton, R., Points, S.D., Smith, R.~C.: 2004, \journal{Astrophys. J}, \vol{613}, 948.

Wright, A.~E., Griffith, M.~R., Burke, B.~F., Ekers, R.~D.: 1994, \journal{Astrophys. J. Suppl. Series}, \vol{91}, 111.

\endreferences

}

\end{multicols}

\vfill\eject

{\ }



\naslov{MULTIFREKVENCIONA POSMATRA{NJ}A NAJVE{\CC}EG OSTATKA SUPERNOVE U LOKALNOJ GRUPI GALAKSIJA -- {\bf LMC SNR J0450--709}} 


\authors{K. O. \v{C}ajko$^{\bf 1}$, E. J. Crawford$^{\bf 2}$ M. D. Filipovi\'c$^{\bf 2}$}

\vskip3mm


\address{$^1$Faculty of Sciences, University of Novi 
Sad\break Trg Dositeja Obradovića 4, 21000 Novi Sad, Serbia} 

\Email{tinacaj@gmail.com}

\address{$^2$School of Computing and Mathematics, University of Western 
Sydney\break Locked Bag 1797, Penrith South DC, NSW 1797, Australia} 

\Email{e.crawford@uws.edu.au, m.filipovic@uws.edu.au}

\vskip3mm


\centerline{\rrm UDK \udc}

\vskip1mm

\centerline{\rit Originalni nauqni rad}

\vskip.7cm

\begin{multicols}{2}

{


\rrm 

U ovoj studiji predstav{lj}amo nove {\rm ATCA} radio-kontinum rezultate posmatra{nj}a ostatka supernove u Velikom Magelanovom Oblaku -- \textrm{\SNR}. Naxli smo da ovaj ostatak supernove ima {lj}uskastu morfologiju koja je ti\-piq\-na za starije ostatke supernovih. Izmerena vrednost radio spektralnog indeksa iznosi \mbox{$\alpha=$--0.43$\pm$0.06}, a dijametra \mbox{{\rm D}$=101.8\times$75.2$\pm$1} parseka. Detektovali smo relativno visok stepen polarizacije gde je maksimalna vrednost iznosi oko 40\%. 

}\end{multicols}

\end{document}